\documentclass[twocolumn]{aastex631}

\shorttitle{}
\shortauthors{Rozner \& Ramirez-Ruiz}

\usepackage{graphicx}	
\usepackage{amsmath}	
\usepackage{amssymb}	
\usepackage{color}
\usepackage{float}
\usepackage{ulem}
\usepackage{soul}
\usepackage{lipsum}
\usepackage[T1]{fontenc}

\DeclareRobustCommand{\VAN}[3]{#2}
\let\VANthebibliography\thebibliography
\def\thebibliography{\DeclareRobustCommand{\VAN}[3]{##3}\VANthebibliography}

\usepackage{graphicx}	
\usepackage{amsmath}	
\usepackage{braket}

\begin{document}

\title{Stellar distributions around supermassive black holes in gas-rich nuclear star clusters}


\email{morozner@ast.cam.ac.uk}

\author[0000-0002-2728-0132]{Mor Rozner}
\affiliation{Institute of Astronomy, University of Cambridge, Madingley Road, Cambridge CB3 0HA, UK}
\affiliation{
Gonville \& Caius College, Trinity Street, Cambridge, CB2 1TA, UK}
\affiliation{
Institute for Advanced Study, Einstein Drive, Princeton, NJ 08540, USA}


\author[0000-0003-2558-3102]{Enrico Ramirez-Ruiz}
\affiliation{Department of Astronomy and Astrophysics, University of California, Santa Cruz, CA 95064, USA}

\begin{abstract} 
 We study the stellar distribution around supermassive black holes (SMBHs) in gas-rich nuclear star clusters (NSCs). NSCs could contain vast amounts of gas, which contribute significantly to shaping the stellar distribution, typically altering the stellar density cusp from the usual \citet{BahcallWolf1976} solution and consequently affecting the dynamics in the NSC. The dense gaseous environment in NSCs gives rise to dynamical phenomena that are otherwise rare in other gas-free environments. Here we extend the derivation introduced in \citet{BahcallWolf1976} to include an additional energy dissipation term associated with gas drag. We examine the effect of different forms of gas drag on the stellar density distribution. Finally, we discuss implications on the rates of tidal disruption events and other transients triggered by stellar interactions in gas-rich galactic nuclei. 
\end{abstract}



\section{Introduction}

Nuclear star clusters (NSCs) are highly abundant dense stellar clusters that can be found in the cores of many galaxies, hosting central supermassive black holes (SMBHs). 
Their high stellar density enables encounters and interactions that are less probable in other environments. 
NSCs constitute a fertile ground for various dynamical phenomena, including gravitational wave (GW) mergers \citep[e.g.,][]{2006ApJ...637..937O,2014ApJ...784...71S,2016PhRvD..93h4029R},  extreme-mass ratio inspirals \citep[e.g.,][]{2007CQGra..24R.113A}, tidal disruption events (TDEs), quasi-periodic eruptions (QPE) and other phenomena \citep[e.g.,][]{1983Ap&SS..95...11Z,Rees1988,1996ApJ...460..207L,1998ApJ...507..131I,2006MmSAI..77..733K,2010MNRAS.402.1614D,GuillochonRamirezRuiz2013, MacLeod2013, MacLeod2014,AmaroSeoane2018,
Gezari2021,2023ApJ...957...34L}.  These stellar encounters motivate a comprehensive study of the properties of NSCs and their associated stellar distribution, which, in turn, determine the rates and characteristics of these events. 
 \citet{Peebles1972} and \cite{BahcallWolf1976,BahcallWolf1977} derived steady-state solutions for the stellar cusp in pioneering works, assuming a power-law ansatz. These solutions were later generalized  \citep[e.g.,][]{LightmanShapiro1977,
 CohnKulsrud1978,
AlexanderHopman2009,Keshet2009,Vasiliev2017,Rom2023}, yet neglecting the effects of circumstellar gas. 

NSCs could acquire significant amounts of gas during their formation or at later stages, when gas is transported and accreted by the central SMBH \citep[e.g.,][]{Naiman2011}. There are two main suggested channels for NSCs formation: in situ (gas-rich formation) and gas-free migration.  
It was suggested that NSCs could form in situ at the nuclei \citep[e.g.,][]{Loose1982,Milosavljevic2004,Bekki2007,Antonini2015}. This formation channel requires fueling the galactic center with available gas content, which could be supplied by gas-rich galactic mergers \citep{MihosHernquist1996}, or magnetic-rotational instabilities in the gas disk \citep{Milosavljevic2004}. Even if NSCs form via migration, they could form as massive clusters in a gas-rich disk, and retain their gas reservoir while migrating towards the center  \citep['wet migration scenario',][]{Guillard2016}.
Gas expulsion in massive star clusters is less efficient, as the binding energy of a cluster scales as its mass squared, and feedback processes scale linearly with the mass \citep{Krause2016}. This allows for the efficient retention of primordial gas even at late stages. 
Another smoking gun for abundant gas content in NSCs is the existence of multiple populations \citep[e.g.,][]{Rossa2006}, which could be explained by gas inflows toward the center of the galaxy. Even the NSC at the Milky Way nuclei shows evidence of a complicated star formation history \citep[e.g.,][]{Do2015}. 

The dynamics of stars in gas-rich environments differ substantially from those in gas-free settings. Gas can serve as a catalyst for stellar interactions that normally occur in gas-free environments as well as giving rise to new phenomena.
The interaction with gas could affect
gravitational wave mergers, TDEs and other transients \citep[e.g.,][]{BahcallOstriker1975,Ostriker1983,Artymowicz1993,
McKernan2012,2015ApJ...802L..22R,Stone2017,Tagawa2020,RoznerPerets2022,KarasSubr2007,Kritos2024,
KaurStone2025,Rozner_mAGN2025,Rowan2025}, as well as the altering the formation rates and properties, and evolution of binaries \citep{Tagawa2020, Dempsey2022,
RoznerGenerozovPerets2023,
Li_etal_2023,Rowan2023,Whitehead2023, Calinco2024,
Dittmann2024,
Dodici2024,
LiLai2024,MishraCalinco2024,
Rowan2024,RoznerPerets2024,Whitehead2024,Whitehead2025}. 

The classical Bahcall-Wolf cusp relates solely to the distribution of the stellar component of the cluster, yet this could be affected by dissipative processes induced by the gaseous environment. In gas-rich clusters, gas could play a role in shaping the stellar distribution and lead it towards a new steady state, or prohibit an equilibrium state to be achieved. These effects could be significant when calculating the rates of various astrophysical phenomena.

In this {\it Letter}, we study stellar distributions in gas-rich nuclear star clusters. We generalize the results of \cite{BahcallWolf1976} to account for the effects of energy dissipation induced by gas, and examine its implications for a wide range of astrophysical phenomena.    
The rest of the paper is organized as follows: in Section~\ref{sec:BW distribution} we review briefly the derivation of Bahcall-Wolf distribution. In section \ref{sec:motion_in_gas} we discuss motion in gas and the scaling of the parameters related to gas in this system. In Section~\ref{sec:BW_gas} we derive the stellar distribution around a SMBH in gas-rich nuclear clusters. In Section~\ref{sec:discussion} we discuss our results and possible caveats. Finally, in Section~\ref{sec:summary} we summarize our findings. 

\section{The Bahcall-Wolf distribution}\label{sec:BW distribution}

The steady-state distribution introduced in \cite{BahcallWolf1976} could be derived from imposing a constant energy flux \citep{ShapiroLightman1976,
BinneyTremaine2008,AlexanderHopman2009}. Assuming the ansatz $n(r)\propto r^p$,

\begin{align}
|\mathcal F_E| \sim N(r) E(r) \Gamma(r)\propto r^{2p+7/2}
\end{align}

\noindent
where $N(r)\propto r^3n(r)$ is the number of stars in radius $r$ from the center of the cluster, $E(r)\propto r^{-1}$ is the energy at radius $r$ and $\Gamma(r)\sim n(r) \sigma v$ is the scattering rate where $n(r)$ is the stellar number density, $\sigma\propto v^{-4}$ is the cross-section, taken to be Rutherford cross-section for small angles and $v$ is the stellar velocity.
$\mathcal{F}_{\rm{E}}$ is the (outward) \textit{energy flux}, i.e., the flux at which energy is carried by stars through energy space at a given specific energy $E$. It quantifies the rate at which energy is transported by interactions -- here, by stellar scatterings -- and later in the letter, we will discuss the energy flux induced by interactions of stars with gas.
Requiring $\mathcal F_E\equiv \rm{constant}$ retrieves the Bahcall-Wolf cusp, i.e. $p=-7/4$.  

\section{Motion in gas}\label{sec:motion_in_gas}

The interaction of stars with gas is complex and involves energy and angular momentum transfer. There are several approaches to model this interaction, including the formation of mini-disks/migration in disks \citep{Artymowicz1991,McKernan2012,Stone2017,Tagawa2020,2023ApJ...944...44K}, following Bondi-like accretion profiles \citep{Antoni2019,2019ApJ...876..142K}, and gas dynamical friction \citep[GDF,][]{Ostriker1999}.
Here we will follow GDF unless stated otherwise, but other models are expected to yield qualitatively similar results. 

The force applied on the object traveling in a gas-rich medium is usually separated into two main regimes, based on the velocity of the object relative to the gas \citep{Ostriker1999}, 

\begin{align}
&|\textbf{F}_{\rm{GDF}}|= \frac{4\pi (Gm)^2\rho_g}{v^2}I(\mathcal M)\hat v; \\
&I(\mathcal M)= \begin{cases}
\frac{1}{2}\ln \frac{1+\mathcal{M}}{1-\mathcal{M}}-\mathcal{M}, \ \mathcal M<1, \\
\frac{1}{2}\ln\frac{\mathcal M+1}{\mathcal M-1
}, \ \mathcal M>1
\end{cases}
\end{align}

\noindent
where $m$ is the mass of the object, $v$ is its velocity, $\rho_g$ is the gas density, $\mathcal{M}=v/c_s$ is the Mach number.
At very low velocities, the force scales linearly with the velocity, and at very high velocities, it scales as $v^{-2}$. Since the sound speed in a NSC is typically much smaller than the stellar velocities, we will mainly focus on the supersonic regime.

Alternatively, being agnostic about the form of the gas drag law, we can parameterize it in terms of the gas density profile and the velocity dependence

\begin{align}\label{eq:gas_drag}
|\textbf{F}_{\rm{gas}}|\propto \rho_g(r) v_{\rm{rel}}^\beta\propto r^\gamma v_{\rm{rel}}^\beta
\end{align}

\subsection{Parameters of the gas}\label{subsec:parameters_of_the_gas}

The gas density in the cluster depends on the mass of the SMBH.
The observed $M-\sigma$ relation suggests a dependence of $M_{\bullet}\propto \sigma^4$, where $\sigma$ is the stellar velocity dispersion \citep{KormendyHo2013}, the dependence of the NSCs masses is shallower, and scales as $\propto \sigma^2$ \citep{ScottGraham2013}. 
We adopt $\sigma= 200 \ \rm{km} \ \rm{sec}^{-1} \left(M_\bullet/10^9 \ M_\odot\right)^{1/4}$ and $M_{\rm{NSC}}=10^7 \ M_\odot \left(M_\bullet/10^7 \ M_\odot \right)^{0.5}$ and set the default SMBH to be $10^8 \ M_\odot$.  
The radius of influence is defined by

\begin{align}
R_{\rm{inf}}= \frac{GM_{\bullet}}{\sigma^2}\propto M_\bullet^{1/2}
\end{align}

\noindent
expressing the gas mass in the cluster in terms of the SMBH, one can estimate

\begin{align}
\rho_{g,0}\sim \xi\frac{M_{\star}}{R_{\rm{inf}}^3}\propto M_\bullet^{-1}
\end{align}

\noindent
We parameterize the contribution of gas by $\xi = M_{\rm{gas}}/M_\star$ where $M_{\rm{gas}}=\rho_{g,0}\int_{R_{\rm{cut}}}^{R_{\rm{inf}}} 4\pi r^2 (r/R_{\rm{inf}})^\gamma dr$ is the mass in gas and $M_\star$ is the total mass in stars in the cluster.

The gas density profile in NSCs is currently uncertain, and several plausible profiles could be considered, usually parameterized as power laws $\rho_g(r)\propto \rho_{g,0}r^\gamma$. At small radii, the profile is expected to follow Bondi accretion profile, i.e. $\gamma=-3/2$, while at larger radii, it is expected to follow a profile dictated by thermally driven wind, i.e. $\gamma=-2$ \citep{Quataert2004,2012ApJ...760..103D}, which corresponds also to an isothermal sphere. See also discussions on plausible gas density profiles in clusters in \citet{LinMurray2007}, \citet{Naiman2009,Naiman2011} and \citet{2019ApJ...876..142K}.

\section{Stellar distribution in gas-rich nuclear star clusters}\label{sec:BW_gas}

Introducing gas to the system modifies the energy flux carried by stars.
This modification could be manifested by adding a drift term to the Fokker-Planck equation that describes the stellar distribution in the cluster, to finally alter the steady state or even bring the system to a state without a steady state. The total energy flux could then be written as 
\begin{align}
&\mathcal{F}_E \sim \mathcal{F}_\star+\mathcal{F}_{\rm{gas}}
\end{align}

\noindent
where the first term is the one discussed in sec. \ref{sec:BW distribution}, and the second one encapsulates the contribution to the energy flux induced by the interaction of stars with the gas.
Assuming a power law ansatz of $n(r)\propto r^p$, the energy flux induced by the general form of gas drag (eq. \ref{eq:gas_drag}) is then given by 

\begin{align}
\mathcal{F}_{\rm{gas}} & \sim N(r) \frac{dE}{dt}\bigg|_{\rm{gas}}\\ \nonumber
&\sim \frac{1}{m}N(r) |\textbf{F}_{\rm{gas}}\cdot \textbf{v}|\propto r^{p+5/2-\beta/2+\gamma}
\end{align}

\subsection{Explicit calculation of the coefficients}\label{sec:explicit_coefficients}
Consider a SMBH with mass $M_{\bullet}$, embedded in a gas-rich NSC that contains stars of equal mass $m$, characterized by \textit{specific} energy $E=GM_\bullet/r-v^2/2$. It should be noted that we adopt the convention used in \cite{BahcallWolf1976}, in which bound orbits carry positive energies, and energies are more positive closer to the SMBH. Unless stated otherwise, we will relate here to the specific energies as {\it energies} for brevity. The distribution function in the phase space $f(E)$ is assumed to follow a power law ansatz, i.e. $f(E)=f_0(E/E_0)^\alpha$, and accordingly the stellar number density could be calculated as $n(r)=\int f(E) d^3 v$. Following \cite{BahcallWolf1976}, we assume the stellar distribution is spherical symmetric in space, isotropic in velocity and governed by weak, slow interactions rather than strong, fast ones.  

The energy flux comprises three terms: advective, conductive and gas-drag. The advective and conductive terms are given by \citep{BahcallWolf1976,Vasiliev2017,Rom2023} 

\begin{align}
&\mathcal F_{\rm{adv}}= D_\star E^{2\alpha-1/2}\frac{3(1/4-\alpha)}{\alpha(\alpha+1)(\alpha-1/2)(\alpha-3/2)},\\
&\mathcal F_{\rm{cond}}= D_\star E^{2\alpha-1/2}\frac{3/2}{\alpha(\alpha+1)(\alpha-1/2)(\alpha-3/2)};\\
&D_\star = \frac{32\sqrt{2}\pi^5}{3}\frac{\alpha f_0^2}{E_0^{2\alpha}}G^5 m^2 M_{\bullet}^3\ln \Lambda_\star
\end{align}

\noindent
where $\ln \Lambda_\star$ is the Coulomb logarithm for stellar interactions, taken to be $10$, and the total energy flux carried by stars due to scattering is given by $\mathcal F_\star=\mathcal F_{\rm{adv}}+\mathcal F_{\rm{cond}}$. 

The number of stars per unit time that are scattered into a region of energy greater than $E$ due to the effect of gas drag (particle flux), i.e., the analog of the flux $R(E)$ introduced in section III in \cite{BahcallWolf1976} is

\begin{align}\label{eq:Rg}
&R_g(E)=\int_{E}^\infty \Gamma_g(E')dE',\\ \nonumber
&\Gamma_g(E)= f(E)g(E)\frac{1}{E}\frac{dE}{dt}\bigg|_{\rm{gas}}
\end{align}

\noindent
where $\Gamma_g(E)$ is the phase-space flux density of stars of energy $E$ induced by gas drag and $g(E)=4\pi^2 (GM_\bullet)^3/|E|^{5/2}$ is the density of states for Keplerian potential. The energy dissipation rate, using GDF in the supersonic regime

\begin{align}\label{eq:dEdtgas}
\frac{dE}{dt}\bigg|_{\rm{gas}}
=+\frac{4\pi G^2 m\rho_g(r) \ln \Lambda_g}{v}  
\end{align}

\noindent
where $\ln \Lambda_g$ is the Coulomb logarithm for gas, taken to be $3.1$ \citep{Tagawa2020}.
It should be noted that the sign is positive to indicate that gas drag brings stars closer to the SMBH, according to the convention of energy becoming more positive towards the SMBH.
The energy flux carried by the stars is then given by 

\begin{align}\label{eq:gas-induced_energy_flux}
&\mathcal F_{\rm{gas}}=ER_g(E)= +D_g\frac{ E^{\alpha-\gamma-2}}{3+\gamma-\alpha}, \\
&D_g = \frac{16\pi^3 G^5 M_\bullet^3m\rho_{g,0}f_0\ln \Lambda_g}{\sqrt{2}E_0^{\alpha-\gamma}}
\end{align}

\noindent
$\alpha-\gamma<3$ is required for the integral convergence.

As can be seen, when including the gas drag term, the Bahcall-Wolf solution is not a steady-state anymore, as expected, but leaves a residual inwards energy flux.

It should be noted that the motion of black holes in the cluster could be subsonic, leading to a different dependence of the energy dissipation on the velocity,

\begin{align}
\frac{dE}{dt}\bigg|_{\rm{gas,sub}}=+\frac{4\pi G^2m \rho_g(r)}{3c_s^3}v^2
\end{align}

\noindent
and accordingly, the gas-induced flux will be

\begin{align}
\mathcal F_{\rm{gas}}=+D_{{\rm{g,sub}}}E^{\alpha-\gamma-1/2}, \\
D_{\rm{g,sub}}=\frac{32\pi^3 G^5 M_\bullet^3\rho_{g,0}f_0}{3c_s^3E_0^{\alpha-\gamma}}
\end{align}

Hereafter, we will focus on \textit{supersonic} motion, unless stated otherwise.

\subsection{Limiting cases}

The gas-induced energy flux perturbs the steady-state introduced in \cite{BahcallWolf1976}. A new steady state can not be achieved generally. We introduce two limiting cases, characterized by the ratio of fluxes $|\mathcal F_{\rm{gas}}/\mathcal{F}_\star|$: scattering-dominated ($|\mathcal F_{\rm{gas}}/\mathcal{F}_\star|\ll1$) and gas-dominated ($|\mathcal F_{\rm{gas}}/\mathcal{F}_\star|\gg1$).
The first one restores the usual Bahcall-Wolf distribution, i.e., $p=-7/4$ or equivalently $\alpha=1/4$, but with residual energy flux towards the center induced by the gas drag. The second case, in which the gas-induced flux dominates, will lead to a steady state distribution of $\alpha=\gamma+2$, as can be seen from setting the gas-induced energy flux in equation \ref{eq:gas-induced_energy_flux} to be constant. 
All the intermediate cases will obey an interpolation between these two limiting cases.  
\subsection{Exploring the parameter space}
Here we will investigate the effect of adding the gas-induced energy flux, and assess its importance in different regions of the parameter space.

\begin{figure}[H]
    \centering
    \includegraphics[width=1\linewidth]
    {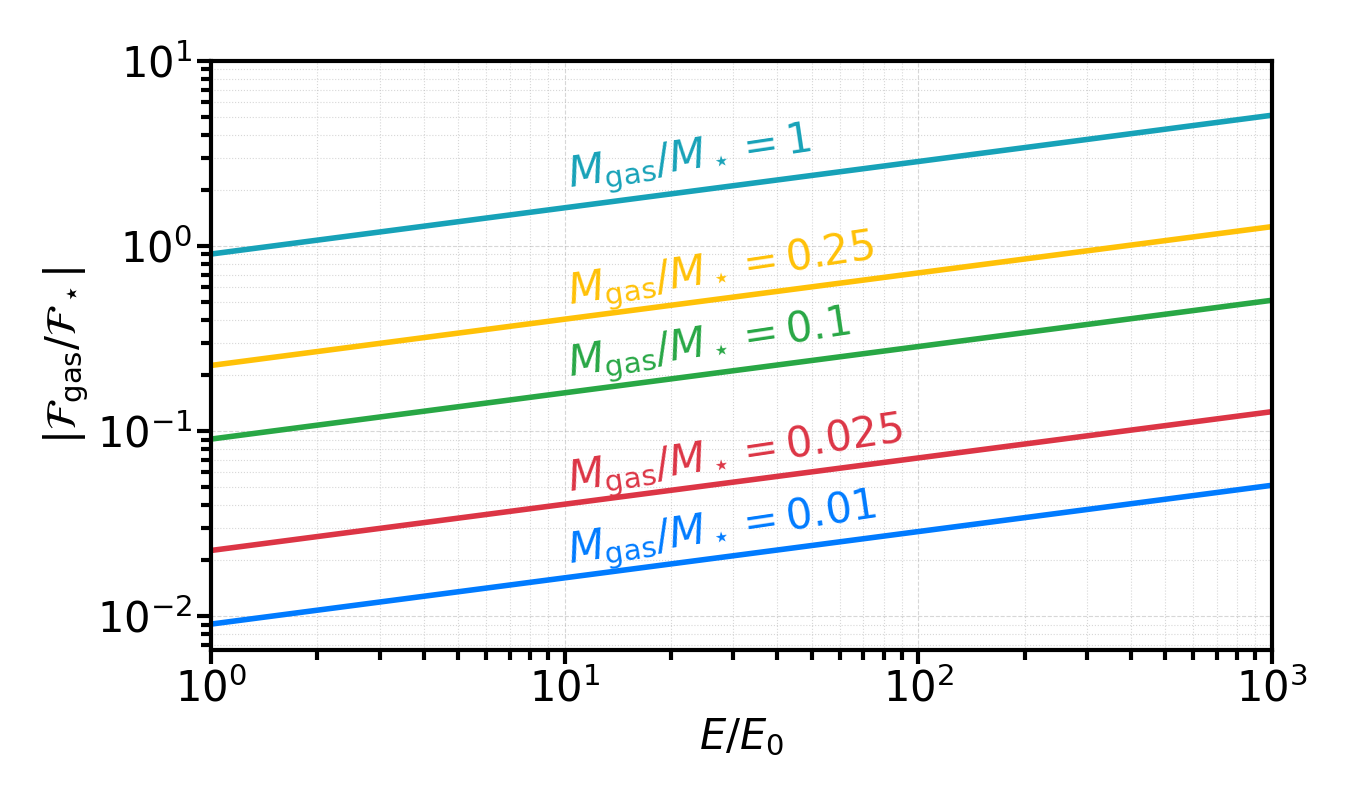}
    \caption{The ratio between gas-induced and scattering energy fluxes, at different locations at the cluster, or equivalently, different specific energies, for different mass ratios between gas and stars. The energies are normalized by the specific energy at the radius of influence $E_0=GM_\bullet/R_{\rm{inf}}$. The power laws of the phase space distribution and gas density are taken to be
    $\alpha=1/4,\gamma=-2$ correspondingly. }
    \label{fig:mass_ratios}
\end{figure}

In Fig. \ref{fig:mass_ratios}, we describe the dependence of the fluxes ratio, $|\mathcal F_{\rm{gas}}/\mathcal{F}_{\star}|$ as a function of the specific energy, for different gas-to-stars mass ratios, considering power laws of $\alpha=1/4$ and $\gamma=-2$ for the phase space distribution and gas density correspondingly. Higher mass ratios/higher energies (or equivalently, closer distances to the SMBH) lead to a larger effect of the gas on the stellar density around the SMBH. Under this choice of parameter, the contribution of gas becomes comparable to the one of stellar-scattering for $M_{\rm{gas}}/M_{\star}\approx 10\%$. It should be noted that the effect of adding gas-induced energy flux is important even when the ratio of fluxes is smaller than one, as it introduces a residual inwards flux that alters the Bahcall-Wolf steady-state distribution.

\begin{figure*}[htbp]
    \centering
    \includegraphics[width=.49\linewidth]{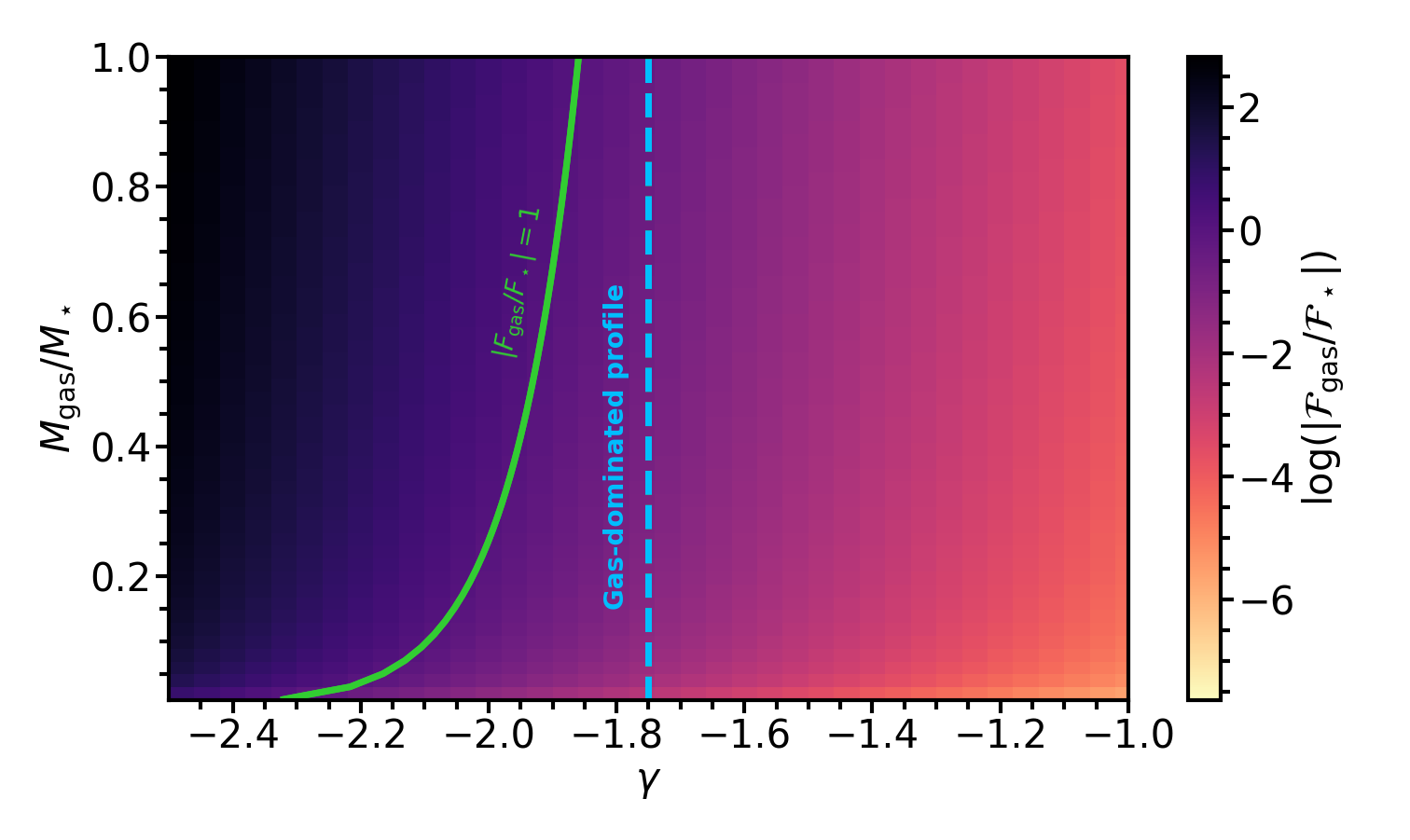} 
    \includegraphics[width=.49\linewidth]
    {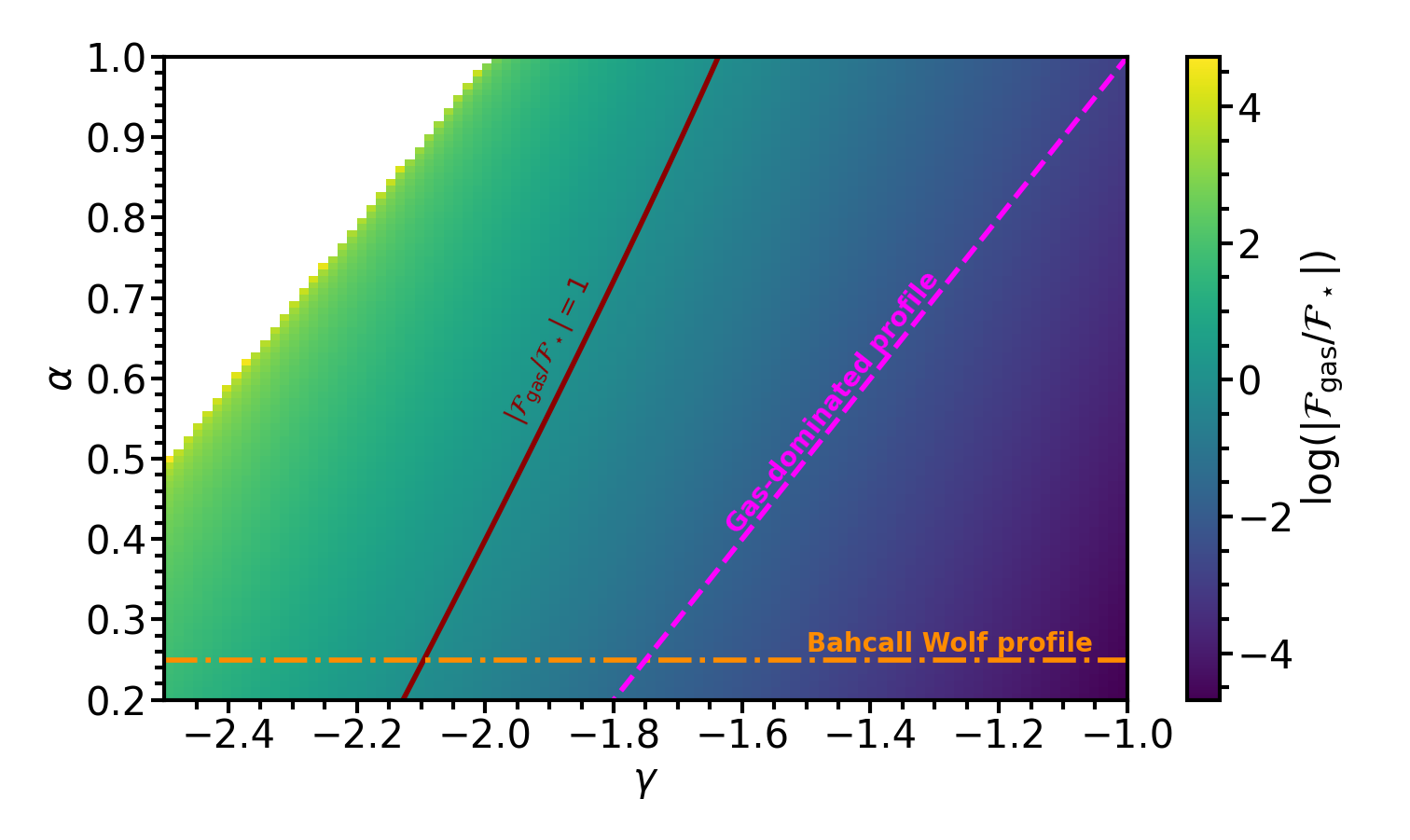}
    \caption{\textit{Left:} Colormap of the ratio of fluxes (color) as calculated for different gas-to-stars mass ratios and gas density power law $\gamma$. The solid line is the contour where $|\mathcal F_{\rm{gas}}/\mathcal F_\star|=1$. The
    energies are evaluated at $0.1 \ \rm{pc}$ and the power law for the phase space distribution is taken to be $\alpha=1/4$.  The dashed line is the profile at the limit of gas-domination, for the given $\alpha$.
    \textit{Right:} Colormap of the ratio of fluxes (color) as calculated for different phase space density power laws $\alpha$ and gas density power laws $\gamma$. The solid line is the contour where $|\mathcal F_{\rm{gas}}/\mathcal F_\star|=1$, the dashed line is the profile at the limit of gas-domination and the dotted-dashed line is the Bahcall-Wolf profile. The energies are evaluated at $0.1 \ \rm{pc}$ and the gas-to-stars mass ratio is set to be $10\%$.
     }
    \label{fig:colormap_gamma}
\end{figure*}

In the left panel of Fig. \ref{fig:colormap_gamma}, we present a colormap of the ratio of fluxes (color) as calculated for different gas-to-stars mass ratios and gas density power law $\gamma$. The solid line is the contour where $|\mathcal F_{\rm{gas}}/\mathcal F_\star|=1$. The
    energies are evaluated at $0.1 \ \rm{pc}$ and the power law for the phase space distribution is taken to be $\alpha=1/4$. The dashed line is the profile at the limit of gas-domination, for the given $\alpha$.
    The contribution from gas-induced energy flux is larger the one from scattering-induced energy flux for parameters above the solid line. 
    As can be seen, steeper gas density slopes correspond to a higher ratio of fluxes. At higher fractions of gas, the contribution from the gas-induced energy flux dominates even for shallower gas profiles. 

In the right panel of Fig. \ref{fig:colormap_gamma}, we present a colormap of the ratio of fluxes (color) as calculated for different phase space density power laws $\alpha$ and gas density power laws $\gamma$. The solid line is the contour where $|\mathcal F_{\rm{gas}}/\mathcal F_\star|=1$, the dashed line is the profile at the limit of gas-domination and the dotted-dashed line is the Bahcall-Wolf profile (stellar-scattering domination). The energies are evaluated at $0.1 \ \rm{pc}$ and the gas-to-stars mass ratio is set to be $10\%$.
As can be seen, steeper gas density and energy power laws, lead to a higher ratio of fluxes. 
The white upper left region stands for a prohibited region in the parameter space, in order to maintain the convergence of the integral in equation \ref{eq:Rg}.

\begin{figure*}[hbtp]
    \centering
    \includegraphics[width=.49\linewidth]{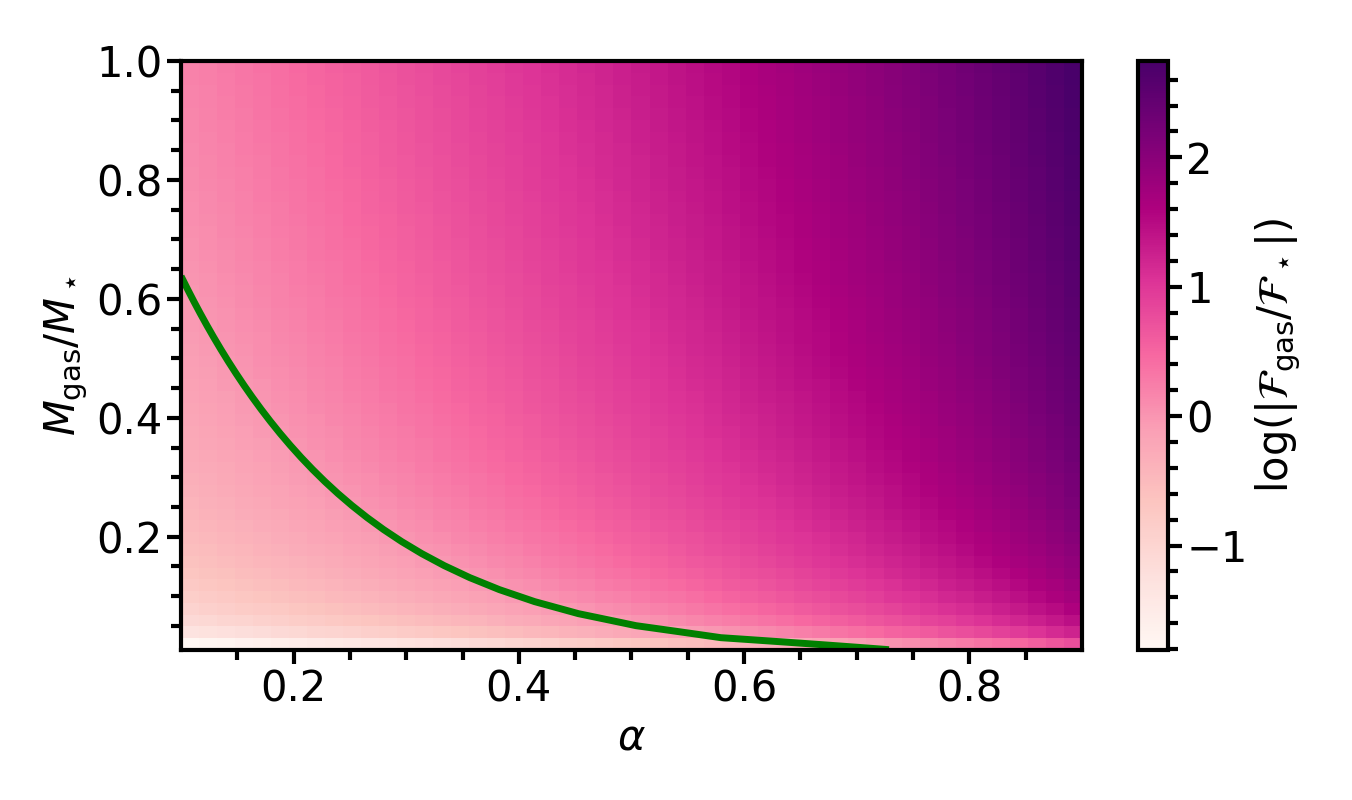}
        \includegraphics[width=.49\linewidth]
    {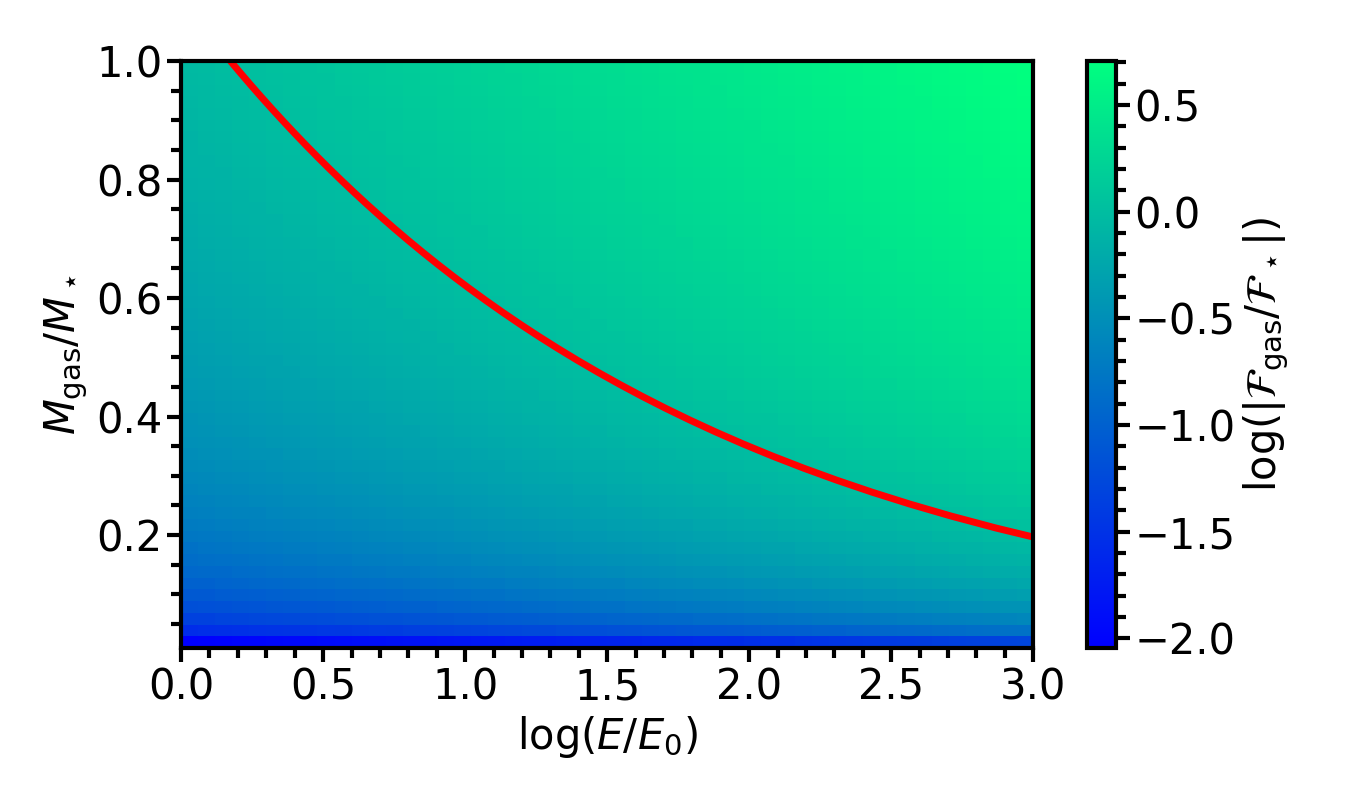}
    \caption{\textit{Left:} Colormap of the ratio of fluxes (color) as calculated for different gas-to-stars ratios and phase space density power laws $\alpha$. The solid line is the contour where $|\mathcal F_{\rm{gas}}/\mathcal F_\star|=1$. The
    energies are evaluated at $0.1 \ \rm{pc}$ and the power law for gas density profile taken to be $\gamma=-2$.
    \textit{Right:} Colormap of the ratio of fluxes (color) as calculated for different gas-to-stars ratios and energies relative to the energy at the radius of influence. The solid line is the contour where $|\mathcal F_{\rm{gas}}/\mathcal F_\star|=1$. The power laws for the phase space distribution and gas density are taken to be $\alpha=1/4,\gamma=-2$ correspondingly.
    correspondingly.}\label{fig:colormap_alpha}
\end{figure*}

In the left panel of Fig. \ref{fig:colormap_alpha}, we present a colormap of the ratio of fluxes (color) as calculated for different gas-to-stars ratios and phase space energy power law $\alpha$.
The solid line is the contour where $|\mathcal F_{\rm{gas}}/\mathcal F_\star|=1$. The
    energies are evaluated at $0.1 \ \rm{pc}$ and the power law for gas density profile taken to be $\gamma=-2$.
    The gas-induced energy flux becomes comparable to the scattering-induced flux
    for parameters above the solid line.
    Smaller values of $\alpha$ require higher fractions of gas for the gas-induced energy flux to be significant. 
In the right panel of Fig. \ref{fig:colormap_alpha}, we present a colormap of the ratio of fluxes for different choices of energies and gas-to-stars ratio. The red line represents the contour where the ratio of fluxes equals $1$, i.e., where the contributions from gas and stars are comparable. Higher energies allow for larger available parameter space in which the gas-induced flux is comparable or larger to the stellar-scattering induced flux.

\section{Discussion}\label{sec:discussion}
The stellar distribution in NSCs plays a role in various dynamical processes, and modifications to it due to interaction with gas could be important for them.
Here we will discuss some of the possible implications.

\subsection{Relaxation \& segregation}\label{subsec:relaxation_segregation}

Stellar relaxation is governed by numerous weak encounters and occasionally strong ones, leading to various dynamical effects such as evaporation, mass segregation, and core collapse. The presence of additional dissipation sources, such as gas, can significantly affect relaxation processes. Gas could accelerate relaxation and mass segregation. The stellar relaxation timescale is given by \citep[e.g.,][]{Spitzer1987} 
$t_{\rm{relax}}\approx 0.065\sigma^3 / G^2 m_\star\rho_\star \ln \Lambda_\star$
where $m_\star$ is the typical stellar mass, $\rho_\star$ is the stellar density and $\ln \Lambda_\star$ is the stellar Coulomb logarithm. Gas introduces another important timescale, $t_{\rm{GDF}}\approx v^3/4\pi G^2 m_\star \rho_g \ln \Lambda_g $ \citep{Ostriker1999}. The added gas-induced flux adds a correction factor to the relaxation time, $\tilde{t}_{\rm{rel}}\approx t_{\rm{rel}}/(1+t_{\rm{rel}}/t_{\rm{GDF}})\approx t_{\rm{rel}}/\left(1+\rho_g/\rho_\star\right)$ where $\rho_g/\rho_\star$ is expected to be of order of few-tens percents \citep[e.g.][]{Guillard2016}.
For massive stars of mass $M$, the segregation time is given by $t_{\rm{seg}}=\braket{m}t_{\rm{relax}}/M$ where $\braket{m}$ is the mean mass of stars in the cluster, such that the segregation is expected to be accelerated by a factor of few. 

\subsection{Enhanced rates of TDEs and other phenomena}
TDEs are thought to be driven by stellar collisions of two-body scattering, characterized by the relaxation timescale. Gas alters the inward flux of stars, allowing them to approach the SMBH at different rates. The typical rate of TDEs is given by 
\begin{align}
\Gamma_{\rm{TDE}}= \frac{N_\star}{\ln\left(J_{\rm{c}}/J_{\rm{lc}}\right)t_{\rm{rel}}}
\end{align}

\noindent
where $N_\star$ is the number of stars contained inside the radius of influence, $J_{\rm{c}}$ is the angular momentum of a circular orbit at the influence radius, $J_{\rm{lc}}$ is the angular momentum for loss-cone orbits and $t_{\rm{rel}}$ is the relaxation time. The relaxation time is modified as we described in the previous subsection (\ref{subsec:relaxation_segregation}), while the relation $J_{\rm{c}}/J_{\rm{lc}}$ roughly doesn't change, as well as $N_\star$. Hence, the typical TDE rate is expected to change by a factor of a few. See also \cite{KaurStone2025} for a detailed discussion on enhanced rates of TDEs in gas-rich AGN disks.

Gas could, in general, drive further stars into the loss cone region, potentially increasing also the rates of stellar collisions \citep[e.g.,][]{2024ApJ...963L..17R}, gravitational waves \citep[e.g.,][]{2014ApJ...784...71S}, and other dynamical phenomena \citep[e.g.,][]{2009ApJ...705L.128R,2010ApJ...720..953L}. Runaway collisions could lead to hierarchical growth of objects and the formation of massive stars and black holes, which are hence expected to be more abundant in gas-rich clusters \citep[e.g.,][]{Vaccaro2024}.
Enhancing the stellar density towards the center also strengthens stellar dynamical friction, inducing a runaway process, that will eventually be halted by feedback.

\subsection{Feedback and gas lifetime} 
The processes described above will be halted eventually, and gas removal could be governed by various processes, including winds, radiation, supernovae, accretion onto the SMBH, and collisions between stars. Indeed, the expected gas lifetime in NSCs is usually estimated by $\approx 10 \ \rm{Myrs}$ \citep[e.g.,][and references therein]{Krause2016}. 

Stars inspiraling towards the center are contributing as well to heating. The binding energy of the gas in the cluster is given by 

\begin{align}
E_{\rm{bind}}= \frac{3GM_{\rm{gas}}^2}{5R_{\rm{cluster}}}+\frac{GM_\bullet M_{\rm{gas}}}{R_{\rm{cluster}}}
\end{align}

Where the first term relates to the self-gravitational binding energy of the gas, and the second one relates to the binding to the SMBH.
Each inspiraling star heats the cluster at a rate of $\dot E|_{\rm{gas}}$, which is specified in equation \ref{eq:dEdtgas}, multiplied by $m_\star$ to convert from specific energy to energy. The characteristic timescale to expel all the gas from the cluster via gas-induced inspiral could be estimated by 

\begin{align}
t_{\rm{expel}}\approx \bigg|\frac{E_{\rm{bind}}}{N_\star \dot E|_{\rm{gas}}}\bigg|\approx 800 \ \rm{Myr} 
\end{align}

\noindent
where we considered here $N_\star$ Solar mass stars, in a cluster with a SMBH mass of $10^6 \ M_\odot$ and gas-to-stars ratio of $10\%$, embedded within the radius of influence. The rest of the parameters follow our description in Section~\ref{subsec:parameters_of_the_gas}. This, as noted, is a generous upper limit for the gas lifetime, which will typically be shorter. 

Gas-rich epochs in NSCs could be periodic rather than a one-time event. If the typical relaxation timescale, which is the timescale in which the stars re-arrange themselves back to the equilibrium gas-free distribution, exceeds the gas-replenishment timescale, the stellar distribution in the cluster could be affected for an extended epoch, longer than the gas lifetime, and oscillate between the gas-dominated distribution and the stellar-scattering dominated one, such that gas-dominated steady-state is not achieved, and the transition between scattering dominated and gas-dominated distributions is determined by the ratio $|\mathcal F_{\rm{gas}}/\mathcal F_\star|$. Similar/shorter timescales are discussed in the context of gas depletion and self-regulation in AGN disks \citep[see e.g.,][and references therein]{Cicone2014,
Trebitsch2019}. 

\subsection{Contributions from the gas potential
}

The contribution from the gas could also alter the potential of the NSC, leading to further corrections to Bahcall-Wolf distribution. The potential dependence in the derivation introduced in subsec. \ref{sec:explicit_coefficients} is encapsulated in the definition of the density of states $g(E)$, and we could extend it for a general potential instead of a Keplerian one. We will demonstrate our derivation for a general power law potential, $\phi(r)\propto r^{-\beta} \ (\beta \neq 0)$. For such a potential, the density of states will be given by $g(E)=dN/dE\propto |E|^{1/2-3/\beta}$. Correspondingly, the modified energy flux induced by gas drag will then scale as $\mathcal F_{\rm{gas}}\propto E^{\alpha-\gamma-3/\beta+1}$.
The energy flux induced by scattering will change as well due to the modified potential. 
Following \cite{Rom2023}, it will scale as $\mathcal F_\star\propto E^{2\alpha-3/\beta+5/2}$. While the dependence of the energy scales the same in terms of $\beta$ for both fluxes, dissimilarities could emerge from the prefactors.

\subsection{Caveats \& future directions}

\begin{itemize}
    \item The motion of objects in gas is not well understood, and several models have been suggested to describe the energy dissipation induced by the interaction with the gas. Here we used GDF \citep{Ostriker1999}, but other models could be implemented, and the qualitative manner of the effect we describe will remain. 
    \item Following \cite{BahcallWolf1976}, we assumed a spherically symmetric spatial distribution, and isotropic velocities. More realistic models should also include angular momentum dependence, and discuss also disk-like geometries, while here we focused on energy fluxes only.
    \item We focused here on a single-mass population, but more realistic clusters, of course, host a rich mass spectrum. The multi-mass stellar distribution derived in \cite{BahcallWolf1977} suggests that the most massive species, $m_{\rm{max}}$, obeys $\alpha_{\rm{max}}=1/4$ energy slope, while lighter species with masses $m_i$ distribute according to $\alpha_i=\alpha_{\rm{max}}m_i/m_{\rm{max}}$, i.e., shallower slopes. To treat this case, one should then carry out the calculation for each species separately. More massive species will be affected more strongly by gas (eq. \ref{eq:gas-induced_energy_flux}), but the contribution from the stellar fluxes will be stronger as well, so the overall direction is not trivial.
    \item 
    Many stars in NSCs could be a part of a binary or higher hierarchy system.
    We ignored their effect on the stellar distribution, and focused on single stars.
    \item A detailed analysis of the problem described here requires solving full Fokker-Planck equations. This will be out of the scope of the letter, and is left for future studies. 
    \item The background stellar density profile is likely not static, and evolves with time when more stars tend to sink to the center. Hence, a more detailed comparison between the contribution of gas and stars should evolve the profiles iteratively, taking into account this feedback rather than comparing to the Bahcall-Wolf steady-state profile.
    \item We assumed here that the gas is stationary, but it could have a non-zero velocity. Qualitatively similar results could be derived if the stellar velocities are dominated by the free-fall velocity of the gas towards the center, $v_{\rm{ff}}=\sqrt{2GM_\bullet/r}$, or in general, for a relative velocity between the stars and the gas that is given by $v_{\rm{rel}}=\sqrt{\sigma^2+v^2_{\rm{ff}}}$. The direction of the gas-induced energy flux will still be directed towards the center as gas drag acts to drain kinetic energy from the stars.
\end{itemize}

\section{Summary}\label{sec:summary}

In this {\it Letter}, we derived analytically the energy flux induced by gas drag acting on the stars in gas-rich NSCs, and its effects on the stellar distribution. 

We showed that the gas-induced energy flux becomes comparable to the scattering-induced flux for $M_{\rm{gas}}/M_\star\gtrsim 10\%$, which is relevant for early stages in cluster evolution. Gas affects the stellar distribution even on smaller gas fractions. The steady-state Bahcall-Wolf distribution will be perturbed by a residual flux of stars towards the center, setting a new cusp distribution.

The stellar distribution around supermassive black holes plays an important role in many astrophysical phenomena. 
We discussed some possible implications for the modified stellar distribution. We considered the enhancement of relaxation and segregation processes, as well as the rates of transients such as tidal disruption events.

Our results suggest a new perspective on the stellar distribution in gas-rich clusters, and could be extended to accommodate different forms of  potentials or varied geometries. The behavior of stellar populations in gas-rich clusters is essentially different, and while interaction with gas has been extensively studied, there are still many unexplored dynamical phenomena.  

\section*{Acknowledgements}
We thank Cathie Clarke and Martin Rees for helpful conversations. Dedicated to the memory of John Bahcall who remains a source of inspiration. ER-R thanks the Institute of Astronomy and the Kavli Institute for Cosmology for hospitality during his Kavli lecture visit where this work was initiated. We thank the Heising-Simons Foundation, NSF (AST-2150255 and AST-2307710), Swift (80NSSC21K1409, 80NSSC19K1391), and Chandra (22-0142) for support.




\bibliographystyle{aasjournal}

\bibliography{example1.bib}





\end{document}